\documentclass[12pt]{article}
\usepackage{amsmath}
\usepackage{amssymb,amsfonts}
\usepackage[all]{xy}
\usepackage{graphicx}
\usepackage[parfill]{parskip}
\usepackage[usenames]{color}

\numberwithin{equation}{section}
\numberwithin{table}{section}

\allowdisplaybreaks

\newcommand{\be}{\begin{equation}}
\newcommand{\ee}{\end{equation}}
\def\beq{\begin{eqnarray}}
\def\eeq{\end{eqnarray}}
\def\ba{\begin{eqnarray}}
\def\ea{\end{eqnarray}}

\def\ep1{\epsilon_1}
\def\eps2{\epsilon_2}
\def\ha{h_0}
\def\hb{h_1}
\def\hc{h_2}
\def\hd{h_3}

\newcommand{\IZ}{\mathbb{Z}}
\newcommand{\IC}{\mathbb{C}}

\newcommand{\IR}{\mathbb{R}}

\newcommand{\IH}{\mathbb{H}}

\newcommand{\im}{\mathrm{Im}}

\newcommand{\Aut}{\mathrm{Aut}}

\newcommand{\nn}{\nonumber}

\newcommand{\cW}{{\cal W}}
\newcommand{\cN}{{\cal N}}
\newcommand{\cM}{{\cal M}}

\newcommand{\cF}{{\cal F}}

\newcommand{\sn}{\mathrm{sn}}

\newcommand{\dudx}{\frac{\partial u}{\partial x}}
\newcommand{\dzdtau}{\frac{\partial z}{\partial \tau}}

\begin{document}

\thispagestyle{empty}

\rightline{\small LPTENS 13/15}

\vskip 3cm
\noindent
{\LARGE \bf Transformations of Spherical Blocks}
\vskip .4cm
\begin{center}
\linethickness{.06cm}
\line(1,0){467}
\end{center}
\vskip .5cm
\noindent
{\large \bf Amir-Kian Kashani-Poor}
 \vskip 0.15cm

\noindent
{\large \bf Jan Troost}
\vskip 0.15cm
{\em \hskip -.05cm Laboratoire de Physique Th\'eorique\footnote{Unit\'e Mixte du CNRS et
    de l'Ecole Normale Sup\'erieure associ\'ee \`a l'Universit\'e Pierre et
    Marie Curie 6, UMR
    8549.}}

  \vskip -0.22cm
{\em \hskip -.05cm Ecole Normale Sup\'erieure}
\vskip -0.22cm
{\em \hskip -.05cm 24 rue Lhomond, 75005 Paris, France}
\vskip 1cm

\vskip0cm

\noindent{\sc Abstract:} We further explore the correspondence between
${\cal N}=2$ supersymmetric $SU(2)$ gauge theory with four flavors on
$\epsilon$-deformed backgrounds and conformal field theory, with an
emphasis on the $\epsilon$-expansion of the partition function natural
from a topological string theory point of view. Solving an appropriate
null vector decoupling equation in the semi-classical limit allows us
to express the instanton partition function as a series in quasi-modular
forms of the group $\Gamma(2)$, with the expected symmetry $W(D_4)
\rtimes S_3$. In the presence of an elementary surface operator,
this symmetry is
 enhanced to an action of $W(D_4^{(1)}) \rtimes S_4$
on the instanton partition function, as we demonstrate via the link
between the null vector decoupling equation and the quantum Painlev\'e
VI equation. \vskip 5cm

\pagebreak

{
\tableofcontents }

\section{Introduction}

The correspondence between $\epsilon$-deformed ${\cal
  N}=2$ supersymmetric gauge theories in four dimensions and
two-dimensional conformal field theories \cite{Alday:2009aq} provides
a fascinating bridge between two highly developed domains of
theoretical physics. The bridge allows for two-way beneficial
exchange. In this paper, in the spirit of
\cite{Fateev:2009aw}\cite{Marshakov:2010fx} and especially
\cite{KashaniPoor:2012wb}, we further explore the ramifications of this correspondence in the case of the gauge theory with four fundamental flavors, which yields the spherical four-point conformal block.

Our main interest on the gauge theory side of the correspondence is
the instanton partition function in an expansion in the topological
string coupling $g_s$ and the deformation parameter $s$. This generalization of the genus
expansion naturally arises upon computation via geometric engineering
and the topological string. In conformal field theory, it corresponds
to
the limit in which both the central charge $c$ and the exchanged
weight $h$ become large, with $h \gg c$. In this limit, it
proves advantageous to express the spherical conformal block in
elliptic variables \cite{Zamo1986}, in particular replacing the cross
ratio $x$ of the four insertion points via a modular parameter
$q_2$. The elliptic recursion relation which the spherical four-point
conformal block satisfies permits a simple proof that the genus
expansion has zero radius of convergence. The main motivations for
studying it nevertheless are threefold: it arises in the
semi-classical limit of the conformal block, the coefficients of the
expansion are quasi-modular forms,\footnote{\label{qm_upto_const}More precisely, the genus expansion of the logarithm of the conformal block yields coefficients that are quasi-modular forms up to the constant (i.e. $q_2$-independent) terms. The latter are furnished by the contribution of the three-point function of Liouville theory to the correlator, identified in \cite{Alday:2009aq} as yielding the perturbative part of the Nekrasov partition function.} and the expansion is natural from
the point of view of topological string theory.
 Understanding the
physical importance of the quasi-modular symmetry from the point of view
of conformal field theory, and drawing lessons for the
non-perturbative formulation of the topological string, remain
important open challenges.

Our path towards the computation of the conformal block in a genus
expansion proceeds via the null vector decoupling equation satisfied
by the four-point correlator with an additional insertion of a
degenerate operator. After the passage to elliptic variables
\cite{Zamo1986}, the computation proceeds in close analogy to our
treatment of $\cN=2^*$ gauge theory in \cite{KashaniPoor:2012wb}. In
particular, the $\tau$-derivatives of the expansion coefficients are
manifestly quasi-modular forms, now of the modular subgroup $\Gamma(2)$ rather than
$\Gamma(1)$, and we again observe experimentally that they also
integrate to quasi-modular forms. The $S$-duality and flavor symmetry
of the $N_f=4$ theory, $W(D_4) \rtimes SL(2,\IZ)$, as identified in
\cite{Seiberg:1994aj}, is manifest in our expressions.\footnote{ This
  symmetry can also be analyzed via Coulomb
  integral formulae in conformal field theory, as  in
  \cite{Giribet:2009hm}, based on results in
  \cite{Zamolodchikov:1986bd}\cite{Fateev:2007qn}.}

The gauge theory interpretation of the four-point conformal block with
an additional degenerate insertion has been proposed in
\cite{Alday:2009fs}: it is conjectured to reproduce the instanton
partition function in the presence of an elementary surface
operator. We will demonstrate that this equation (before taking the
semi-classical limit) essentially coincides with the quantum
Painlev\'e VI equation
\cite{LO}\cite{Nagoya}\cite{Zabrodin:2011fk}. The Painlev\'e VI
equation is the most general (rational) ordinary second order
differential equation with no movable singularities except poles
\cite{P}\cite{F}\cite{G}. Its quantization can be performed by reformulating
the equation in terms of a first order Hamiltonian system, then
quantizing the Hamiltonian and studying the corresponding
Schr\"odinger equation. The symmetries of the resulting equation have been
extensively studied in \cite{Nagoya:2012tv}. By mapping this
discussion to our variables, we deduce the action of the extended
symmetry group $W(D_4^{(1)}) \rtimes S_4$ on the instanton partition
function in the presence of the surface operator.

The organization of this paper is as follows. In section \ref{review},
we review the spherical four-point conformal block and the natural
elliptic variables corresponding to the problem. On the gauge theory
side, we review how the symmetry group $W(D_4) \rtimes SL(2,\IZ)$
arises in $N_f=4$ Seiberg-Witten theory. Section \ref{nvd} identifies
the null vector decoupling equation, which governs the instanton
partition function in the presence of an elementary surface operator,
with the quantized Painlev\'e VI equation and discusses the symmetry
group $W(D_4^{(1)}) \rtimes S_4$ of this equation. We then review the
mapping of this equation to elliptic variables, with many details of
the derivation provided in appendix \ref{ellipticvariables}. The
semi-classical limit of the null vector decoupling equation is treated
in section \ref{gen_exp}. There, we discuss the proof of
non-convergence of the genus expansion based on the elliptic recursion
relation for the conformal block. We also demonstrate that under the
assumption of quasi-modularity, this relation 
allows for a computation of
the corresponding topological string amplitudes in the field theory
limit. Finally, we turn to the solution of the null vector decoupling
equation in the semi-classical limit, which intrinsically gives rise to
quasi-modular results for these amplitudes in the
$\epsilon_2 \rightarrow 0$ limit.

\section{The two sides of the correspondence}
\label{review}
In this section, we will review the four-point 
spherical conformal block and $\epsilon$-deformed ${\cal N}=2$ supersymmetric Yang-Mills
theory with $SU(2)$ gauge group and four flavors. The correspondence \cite{Alday:2009aq} equates the former with the instanton partition function of the latter.

\subsection{The four-point conformal block on the sphere} \label{intro_4pt}
In this subsection, we wish to review the notion of the four-point
conformal block on the sphere, and explain how elliptic variables arise naturally in this context.
We consider the four-point correlator 
\be 
C_4  = \langle V_{h_1}(z_1) V_{h_2}(z_2) V_{h_3}(z_3) V_{h_4}(z_4) \rangle 
\ee 
on the sphere of Virasoro primary fields
$V_{h_i}(z_i)$ of weight $h_i$. The cross ratio of the four insertion points,
\be 
x = \frac{(z_3-z_4)(z_2-z_1)}{(z_3-z_1)(z_2-z_4)} \,, 
\ee
 is a
conformal invariant that serves as a
coordinate on the complex structure moduli space $\cM_{4,0}$ of the
four-punctured sphere. The conformal transformation
 \be z' =
\frac{(z-z_4)(z_2-z_1)}{(z-z_1)(z_2-z_4)} 
\label{coordinatetransformation}
 \ee
maps the points $(z_1,z_2,z_3,z_4)$ to the points $(\infty, 1,x,0)$. Conformal invariance determines the $z_i$ dependence of the four-point correlator $C_4$ up to a function of the cross ratio,
\ba
C_4 &=& \left| \frac{(z_4-z_1)(z_2-z_1)}{z_2-z_4} \right|^{2\sum h_i} \prod_{i \neq 1} |z_i-z_1|^{-4h_i} \lim_{z \rightarrow z_1} |z-z_1|^{-4 h_1} \langle V_{h_1}(z'(z))  V_{h_2}(1) V_{h_3}(x) V_{h_4}(0) \rangle \nn \\
&=&\left| \frac{(z_4-z_1)(z_2-z_1)}{z_2-z_4} \right|^{2\sum h_i-4h_1} \prod_{i \neq 1} |z_i-z_1|^{-4h_i} \,G_{1234}(x)  \,. 
\ea
We have here introduced the matrix element $G_{1234}$ via
\be
G_{1234}(x) = \lim_{z \rightarrow z_1} |(z'(z)|^{4h_1} \langle V_{h_1}(z'(z))  V_{h_2}(1) V_{h_3}(x) V_{h_4}(0) \rangle = \langle h_{1} | V_{h_2}(1) V_{h_3} (x) | h_4 \rangle \,,
\ee
with $|h_4 \rangle$ and $\langle h_1|$ defined with regard to the $z'$ coordinate system. The matrix element $G_{1234}$ in turn can be expressed as 
\be
G_{1234}(x) = \sum_h C_{12 h} C_{34}^{h}  \cF_{14}^{23}(h|x)
\bar{\cF}_{14}^{23}(h|\bar{x}) \label{intro_conf_block} \ee
 in terms
of the conformal blocks $\cF_{14}^{23}(h|x)$. By expressing $G_{1234}(x)$ in this fashion, we have separated the contributions which are determined completely by
the Virasoro algebra, encoded in the conformal blocks, from the
content specific to a given conformal field theory, encoded in the
three point functions $C_{ij}^k$. The sum is over the conformal
families that appear both in the operator product expansion of $V_{h_3}$
and $V_{h_4}$, and of $V_{h_1}$ and $V_{h_2}$.
We call these conformal families intermediate, with an associated 
intermediate highest weight.

The matrix element $G_{1234}$ is related to the four-point correlator with insertions at positions $z_i$ via the coordinate transformation (\ref{coordinatetransformation}). 
A different choice of coordinate transformation permutes the
assignment of the points $z_i$ to the points $\{0,1,\infty\}$.  It
acts on the cross ratio $x$ via a rational transformation, and
establishes equalities (so-called channel dualities) between the various
functions $G_{ijkl}(x)$
with permutations of the indices and rational transformations of the
argument $x$, as follows:
\be
G_{1234}(x) = G_{1432}(1-x) = \frac{1}{|x|^{4h_3}} \, G_{4231}(1/x) \,.
\ee
Via equation (\ref{intro_conf_block}), this yields
identities for conformal blocks, {\it after} summation over the respective
intermediate momenta.

The conformal blocks are functions of the cross ratio $x$, which serves as a representative of the 4-tuple $(z_1,z_2,z_3,z_4)$ under the equivalence relation of conformal transformations. We can associate
an elliptic curve to the equivalence class of this tuple via the
equation \be y^2 - t(t-1)(t-x) = 0 \,, \ee with 2-torsion points lying
at $(t,y)=(0,0),(1,0),(x,0)$.\footnote{To see this, transform this
  equation into Weierstrass form by eliminating the quadratic term in
  $t$ via the substitution $t' = t - \frac{x+1}{3}$ and correctly
  normalizing the cubic term via $y' = 2 y$. The equation is hence
  solved by $(t',y') = (\wp(z),\wp'(z))$, and the zeros of $\wp'(z)$
  clearly lie at $\wp(z)= - \frac{x+1}{3}, 1- \frac{x+1}{3}, x-
  \frac{x+1}{3}$.} 
Choosing the first two as a basis in the group of
2-torsion points, this defines a
map from equivalence classes of four insertion points on a sphere to
the equivalence classes of elliptic curves together with a basis of
their group of 2-torsion points. It is not hard to show that the
latter is isomorphic to $\IH/\Gamma(2)$,  the upper half plane $\IH$ modded out by the
group $\Gamma(2)$ of matrices in $SL(2,\mathbb{Z})$ which are equal to the identity
modulo $2$.\footnote{$\IH/SL(2,\IZ)$
  describes equivalence classes of elliptic curves. Modding out by the
  smaller group $\Gamma(2)$ avoids identifying tori whose 2-torsion
  points are permuted.} 
Conversely, given an elliptic curve with modular parameter $\tau$ together
with such a basis, we can define a map to the class of four tuples via
\be (z_1,z_2,z_3,z_4) =
(\infty,\wp(\tfrac{1}{2}),\wp(\tfrac{\tau}{2}+\tfrac{1}{2}),\wp(\tfrac{\tau}{2})) 
\ee 
by invoking the Weierstrass $\wp$-function.
The invariant cross ratio of this tuple is
\be \label{x_of_tau}
x = \frac{e_3-e_2}{e_3-e_1} \,, 
\ee 
where we have introduced the standard nomenclature for the
half-periods of $\wp$, see equation (\ref{half_periods}). The cross
ratio $x$ is invariant under the action of $\Gamma(2)$ on $\tau$. We
have thus established the isomorphism between classes of 4-tuples on
the sphere and $\IH/\Gamma(2)$. We can hence equally well express the
conformal blocks $\cF_{14}^{23}(h|x)$ as functions of $[\tau] \in
\IH/\Gamma(2)$. The permutation of the points $z_i$, which acts as a
rational transformation on the cross ratio $x$, acts via an element of
$SL(2,\IZ)/\Gamma(2)$ on $\tau$. The action of $\Gamma(2)$ on $\tau$
gives rise to additional symmetries not accessible when the blocks are
expressed as functions of $x$. These will play a central role in
section \ref{gen_exp}.

When treating the five
point function on the sphere in the next section, it will prove
convenient to extend this isomorphism to a (necessarily multi-valued)
map from the sphere to the torus which maps the four insertion points
$(\infty, 1,x,0)$ on one sheet to the 2-torsion points. The 
fifth insertion point is then parameterized by an elliptic variable.

\subsection{The ${\cal N}=2$ supersymmetric $SU(2)$ theory with four flavors} 

\label{rev_nf4}
On the four-dimensional side of the two-dimensional/four-dimensional
correspondence, we mostly consider a gauge theory observable that has
several incarnations. It was introduced in the course of the
computation of the prepotential of $\cN=2$ gauge theories via
instanton calculus. As the integrals that arise in this computation
are divergent, one can consider their equivariant counterpart
$Z_{inst}(\epsilon_1,\epsilon_2)$, introducing the two parameters
$\epsilon_1$ and $\epsilon_2$ \cite{Nekrasov:2002qd}. In this
incarnation, the partition function $Z_{inst}$ is defined as a formal
sum in the instanton counting parameter $x=q_{UV}$ corresponding to the
$UV$ coupling. The partition sum $Z_{inst}$ can also be obtained from
the topological string partition function with the appropriate target
space geometry to engineer the field theory. Via the holomorphic
anomaly equations, this definition gives rise to all order results in
the instanton counting parameter, order by order in the
topological string coupling squared $g_s^2 = \epsilon_1 \epsilon_2$
and the deformation parameter $s=(\epsilon_1 + \epsilon_2)^2$.

We concentrate on the $SU(2)$ gauge theory with $\cN=2$ supersymmetry
and $N_f=4$ flavors,
 which is superconformal in the massless limit. The
space of marginal deformations of the theory can be parameterized by a
parameter $\tau$ taking values in the upper half plane. The flavor symmetry is enhanced from $SU(4)$ to $SO(8)$. Since this feature is
important to us, we will review how it arises.

 An $\cN=2$ theory
with $N_f$ massless hypermultiplets $(Q_i,\tilde Q_i)$, such that $Q_i$
transform in the fundamental and $\tilde Q_i$ in the anti-fundamental
of the gauge group, generically enjoys $SU(N_f)$ flavor symmetry. The flavor symmetry is enhanced to $SO(2N_f)$ when the hypermultiplets transform in a pseudo-real representation of the gauge group, such as the fundamental representation for the gauge group $SU(2)$. A pseudo-real representation exhibits an antisymmetric intertwiner $R$ which relates the generators $\sigma_a$ within the representation to their complex conjugates, via $\sigma_a^\ast = - R^{-1} \sigma_a R$. The kinetic
terms of the theory therefore exhibit $SU(2N_f)$ symmetry, made manifest by introducing the field
$\tilde Q_f = R \tilde Q$. The
${\cal N}=2$ superpotential term $\tilde Q \Phi Q$, with $\Phi$ the chiral
superfield in the $\cN=2$ vector multiplet, can be rewritten as
\be
(Q^T \, \tilde Q_f^T)
\begin{pmatrix}
  0 & 1\\
  1 & 0 \\
\end{pmatrix}
\Phi R^{-1} 
\begin{pmatrix}
  Q \\
  \tilde Q_f\\
\end{pmatrix} \,.  
\ee 
This term hence breaks the flavor symmetry from $SU(2N_f)$ to $SO(2N_f)$. Note that a real representation, i.e. a symmetric intertwiner
$R$, would have required the matrix $\left( \begin{smallmatrix} 0 & 1
    \\ -1 & 0 \end{smallmatrix} \right)$ instead, breaking the flavor
symmetry to $Sp(2N_f)$. Returning to the gauge group $SU(2)$ with $N_f=4$ fundamental flavors, the flavor symmetry is thus enhanced from $SU(4)$ to $SO(8)$, as announced.

A mass term consistent with $\cN=2$
supersymmetry is provided by the superpotential term 
\be
 m_i \tilde Q_i
Q_i= (Q_i^T \, \tilde Q_{f,i}^T)
\begin{pmatrix}
  0 & m_i\\
  -m_i & 0 \\
\end{pmatrix}
R^{-1} 
\begin{pmatrix}
  Q_i \\
  \tilde Q_{f,i}\\
\end{pmatrix} \,.  
\ee
Note that the mass matrix takes values in a Cartan of $\mathfrak{so}(8)=D_4$. We can expand it in terms of the standard basis $\{H_i \}$ for the
Cartan subalgebra as $M= \sum_i m_i
H_i$. The mass matrix breaks the flavor symmetry to $SO(2)^4$.  The Weyl
group $W(D_4)$ of the $D_4$ Lie algebra can be identified as the subgroup of $SO(8)$ which maps the Cartan of $D_4$ to itself under conjugation. The vector $(m_1, \ldots, m_4)$ hence furnishes a representation of this group, which acts on it via permutations and an even number of sign changes. This action on the parameters $m_i$ is therefore a symmetry of the theory.

The maximal regular subalgebra $SU(2)^4
\subset SO(8)$ will play a distinguished role in the following. Given
the basis vectors $\epsilon_i$ of root space dual to the generators
$H_i$ of the Cartan introduced above, a convenient choice for the four
roots of $D_4$ corresponding to these four $\mathfrak{su}(2)$ subalgebras is
\be \label{affine_dynkin} \alpha_0 =
-\epsilon_1-\epsilon_2 \,, \quad \alpha_1=\epsilon_1-\epsilon_2
\,, \quad \alpha_3= \epsilon_3-\epsilon_4 \,, \quad \alpha_4=
\epsilon_3+\epsilon_4 \,.  
\ee
We can express the mass matrix $M$
in terms of the basis $E_i^3=  
\alpha_i \cdot H$, $i=0,1,3,4$, of the Cartan by
introducing the mass parameters $M_i$ as follows:
\be
M_0 = -\frac{m_1 + m_2}{2} \,, \quad
M_1 = \frac{m_1-m_2}{2} \,,
\quad
 M_3= \frac{m_3-m_4}{2} \,,
\quad
M_2 = \frac{m_3 +m_4}{2} \,.
\ee
The $N_f=4$ theory also exhibits a strong/weak coupling duality. The
duality group acts via $SL(2,\IZ)$ on the complexified gauge coupling
$\tau$, which can be identified with the argument of the half-periods
$e_i$ in the definition of the original Seiberg-Witten curve
\cite{Seiberg:1994aj}. Simultaneously, it acts on the spectrum of the theory by permuting fundamental representations of the flavor symmetry $SO(8)$
\cite{Seiberg:1994aj}. The full
duality group of the theory
 is hence $W(D_4) \rtimes SL(2,\IZ) $. The subgroup
$\Gamma(2) \subset SL(2,\IZ)$ is a symmetry of the theory acting
solely on the parameter $\tau$. Replacing $\tau$ by the parameter $x=
\frac{e_3-e_2}{e_1-e_2}(\tau)$ makes this invariance manifest. The remaining symmetry group $W(D_4) \rtimes S_3$ is isomorphic to the Weyl group of the exceptional group $F_4$. Under
$S_3 \cong SL(2,\IZ) / \Gamma(2)$, the ratio $x$ transforms as a cross ratio and
the representations of $SO(8)$ are exchanged by triality. The precise
implementation of the semi-direct product $W(D_4) \rtimes SL(2,\IZ) $ on the coupling and the Lie
algebra is reviewed in appendix \ref{group}.

The mass parameters $m_i$ (or $M_i$) can be organized in terms of the $SO(8)$ invariants 
\begin{eqnarray}
R &=&  \frac{1}{2} \sum_i m_i^2 \,, \nonumber \\
T_1 &=& \frac{1}{12} \sum_{i<j} m_i^2 m_j^2 - \frac{1}{24} \sum_i m_i^4 \,,
\nonumber \\
T_2 &=& -\frac{1}{24} \sum_{i<j} m_i^2 m_j^2 + \frac{1}{48} \sum_i m_i^4 - \frac{1}{2} \prod_i m_i \,,
\nonumber \\
T_3 &=& -\frac{1}{24} \sum_{i<j} m_i^2 m_j^2 + \frac{1}{48} \sum_i m_i^4 + \frac{1}{2} \prod_i m_i \,,
\nonumber \\
N &=& \frac{3}{16} \sum_{i<j<k} m_i^2 m_j^2 m_k^2 - \frac{1}{96} \sum_{i \neq j} m_i^2 m_j^4 + \frac{1}{96} \sum_i m_i^6 \,,  \label{casimirs}
\end{eqnarray}
such that $R$ and $N$ are invariant under triality, while the $T_i$ are permuted \cite{Seiberg:1994aj}.

\subsection{The parameter map}
The mapping of parameters that identifies the four-point conformal block $\cF_{14}^{23}$ 
of  a conformal field theory with central charge $c$, insertions of conformal dimension
$h_i$ and exchanged dimension $h$
with the instanton partition function $Z_{inst}$ of the gauge theory
with deformation parameters $\epsilon_i$,
masses $M_i$ and vector adjoint scalar vacuum expectation value $a$ is \cite{Alday:2009aq}
\be  \label{dictionary}
c = 1 + 6 Q^2
\, , 
\quad
Q = b+b^{-1}
\, , 
\quad
b= \sqrt{\frac{\epsilon_2}{\epsilon_1}} \,, 
\quad 
h_i = \frac{Q^2}{4} - \frac{M_i^2}{\epsilon_1 \epsilon_2} \,, 
\quad 
h = \frac{Q^2}{4} - \frac{a^2}{\epsilon_1 \epsilon_2} \,,
\ee
or equivalently, in terms of the exponents $\alpha_i$ of Liouville vertex operators,
\begin{equation}
\alpha_i = \frac{Q}{2} - \frac{M_i}{\sqrt{\epsilon_1 \epsilon_2}} \,, 
\quad
 \alpha = \frac{Q}{2} - \frac{a}{\sqrt{\epsilon_1 \epsilon_2}} \,.
\end{equation}
Note that the massless limit $M_i=0$ corresponds to the insertion of four puncture operators of weight $\frac{Q^2}{4}$.

The $U(1)$ correction factor discussed in \cite{Alday:2009aq}
contributes $a$ independent terms to the leading terms $F^{(n,g)}$,
$n+g \le 1$, of the genus expansion of the partition function that we
will analyze at length. These contributions will not be relevant to our
discussion.

\section{The null vector decoupling equation}
\label{nvd}
To compute the four-point conformal block, we wish to pursue the same strategy 
we employed in \cite{KashaniPoor:2012wb} in the case of the one-point toroidal block: the correlator with a fifth insertion, chosen to
be degenerate, satisfies a null vector decoupling equation. In fact,
each individual conformal block contributing to the correlator satisfies the
equation. Imposing appropriate monodromy conditions
selects the solution coinciding with the exchange of a given
momentum. In the semi-classical limit, we can extract
the semi-classical four-point conformal block from this result.

The five-point function with one degenerate insertion itself has a
conjectured gauge theory interpretation. Indeed, the paper \cite{Alday:2009fs}
proposes
 to identify the corresponding five-point conformal block with
the instanton partition function in the gauge theory in the presence
of a surface operator. According to this proposal, the two integers
which label degenerate operators, see (\ref{deg_weights}),
 determine the
location of the surface operator in $\IR^4$, while the position of the
degenerate operator on the Riemann surface maps to two real parameters
determining the type of surface operator.

It is hence worthwhile to discuss the null vector decoupling equation
the five-point correlator satisfies in generality in this section, before solving it in the
semi-classical limit in the next. We will begin by reviewing how this
equation has arisen in the mathematical physics literature as a
quantization of the Painlev\'e VI equation. By matching conformal
field theory conventions to those employed in \cite{Nagoya:2012tv}, we
can translate the results presented there on the symmetries of the
quantum Painlev\'e equation into our variables. Aside from recovering
the gauge theory symmetries discussed in section \ref{rev_nf4}, 
we
will witness an $\epsilon$-dependent deformation of one of these symmetries,
as well as an enhancement,
 in the case of the
instanton partition function in the presence of a surface
operator. Finally, we will review the mapping of the
null vector decoupling equation into elliptic variables, the starting
point for our semi-classical treatment in section \ref{gen_exp}.

\subsection{The null vector decoupling equation in spherical coordinates}
We consider four primary vertex operators $V_{h_i}$ of weight $h_i$ in
a conformal field theory on the sphere, or the infinite plane,
inserted at points $z_i$. To this four-point correlator, we add a
fifth, degenerate field, which is to stay light in the semi-classical
limit $b \rightarrow 0$. The degenerate weight
\be \label{deg_weights} h_{mn} = \frac{Q^2}{4} - \frac{1}{4} (m b + n
b^{-1})^2 \ee 
of lowest degree which remains light in the semi-classical limit is carried by the primary field
$V_{(2,1)}$ of conformal dimension $h_{2,1}=- \frac{1}{2}-\frac{3}{4} b^2$.
Imposing decoupling of this Virasoro null vector leads to a second order differential
equation on the five-point function,
\be
\left[ \partial_z^2 +b^2 \left( \sum_{k=0}^3 \frac{h_k}{(z-z_k)^2} + \frac{ \partial_k}{z-z_k} \right) \right] \langle V_{(2,1)}(z) V_{\ha} (z_0) \ldots V_{\hd}(z_3) \rangle = 0 \,.
\ee
Mapping the points $(z_0,z_1,z_2,z_3)$ to $(\infty, 1,x,0)$,  the null vector decoupling equation
on the five-point correlator assumes the form
\begin{eqnarray}
0 & = & \Bigg[\frac{1}{b^2} \partial_z^2 + 
  \frac{2 \ha}{z(z-1)} + \frac{\hb}{(z-1)^2}+\frac{\hc}{(z-x)^2} + \frac{\hd}{z^2}
\nonumber \\
 & & - \frac{1}{z (z-1)(z-x)} \Big[ (z-x)  \sum_{i=0}^3 h_i - (x-1) x \partial_{x}  \Big] 
\nonumber \\
& & \label{null_vect_one}
+ \frac{1}{z(z-1)} \Big[ -h_{2,1} + (1-2z) \partial_z \Big] \Bigg]
\Psi_5(z,x) \, .
\end{eqnarray}
with
\be
\Psi_5(z,x) = \langle  V_{\ha}(\infty) V_{(2,1)}(z) V_{\hb}(1) V_{\hc}(x) V_{\hd}(0) \rangle  \,.  
\ee

\subsection{Symmetries of the null vector decoupling equation}
\label{quantumsymmetries}
In the present subsection, we will relate the null
vector decoupling equation (\ref{null_vect_one}) 
to the quantized Painlev\'e VI equation. This will allow us to exploit the results of \cite{Nagoya:2012tv} to elucidate the symmetry group of the null vector
decoupling equation
 and the transformation properties of the
five-point correlator in the two-dimensional theory. The
correspondence \cite{Alday:2009fs} then yields a prediction for the
transformation of the vacuum expectation value of a surface operator
in the $\epsilon$-deformed four-dimensional gauge theory under
strong-weak coupling duality.  The most interesting transformation
rule is non-local, implemented via a Laplace transform.
  We stress that the discussion in
this subsection is valid at finite $\epsilon_1$ and $\epsilon_2$.

We begin by reviewing how the null vector decoupling equation has arisen in the mathematics literature in the context of the theory of differential
equations and their quantization.  The study of (rational) second
order ordinary differential equations which have solutions of which the
movable singularities are necessarily poles \cite{P}, led to a
classification of such equations \cite{G}. The complete set contains six equations, referred to as the Painlev\'e equations I through VI.
 The Painlev\'e VI equation is the
most general one from which all others can be obtained through
degeneration limits. This equation also arises as the differential
equation satisfied by the singularities of a second order Fuchsian
equation with four regular singular points on the two-dimensional
sphere, when one imposes the property that the monodromy of the
solution be preserved \cite{F}. The analysis of isomonodromic
deformations \cite{S}\cite{Garn1} of differential equations leads to a
rich connection \cite{Garn2} with the theory of integrable systems
(see e.g. \cite{Jimbo:1978nt}). The second order Painlev\'e differential equations can be 
re-expressed as a first order Hamiltonian system with interesting algebraic geometric implications \cite{O}.
More recently, the Hamiltonian
Painlev\'e systems have been quantized \cite{Nagoya}, preserving the full symmetry
of the classical problem \cite{Nagoya:2012wm}. The resulting Schr\"odinger equation is essentially the null vector decoupling equation
 (\ref{null_vect_one}).\footnote{Concretely, the quantum
Painlev\'e Schr\"odinger equation in the symmetrized form of equation (2.1) in
\cite{Nagoya:2012tv} acting on a wave-function $\Psi_{sym}$ is
equivalent to the null vector decoupling equation acting on our
five-point correlator $\Psi_5$ on the sphere under the following identification
of parameters (to avoid an internal clash of notation, the
  quantities $\beta_i$ refer to the quantities denoted $\alpha_i$ in
  \cite{Nagoya:2012tv} and it should be understood that
$(q,z)$ in \cite{Nagoya:2012tv} are denoted
  $(z,x)$ here):
\begin{equation}
\beta_1 = \frac{2 M_0}{\epsilon_1} \,,
\quad
\beta_3 = \frac{2 M_1}{\epsilon_1} \,,
\quad
\beta_0 = \frac{2 M_2}{\epsilon_1} \,,
\quad
\beta_4 = \frac{2 M_3}{\epsilon_1} \,,
\quad
\beta_2 = \frac{1}{2} \frac{\epsilon_2}{\epsilon_1}
- \frac{\sum_{i=0}^3 M_i}{\epsilon_1} \,,\quad \kappa = -b^2 \,, 
\end{equation}
and
\begin{eqnarray}
\Psi_5 &=& 
(1 - x)^{- \frac{1}{6b^2} - \frac{1}{2}-\frac{b^2}{4}+ \frac{1}{\epsilon_1 \epsilon_2} (M_1^2+M_2^2)}
 x^{- \frac{1}{6b^2} - \frac{1}{2}-\frac{b^2}{4}+ \frac{1}{\epsilon_1 \epsilon_2} (M_2^2+M_3^2)} 
\nonumber \\
& & 
(x - z)^{\frac{1}{2}+\frac{b^2}{2}-\frac{M_2}{\epsilon_1}} 
(z-1)^{\frac{1}{2}+\frac{b^2}{2}-\frac{M_1}{\epsilon_1}} z^{\frac{1}{2}+\frac{b^2}{2}-\frac{M_3}{\epsilon_1}}
\Psi_{sym} \, .
\end{eqnarray}}

In \cite{Nagoya:2012tv}, it was demonstrated that the full symmetry
group of the Painlev\'e VI equation, determined to be the affine Weyl
group $W(F_4^{(1)})$ of the exceptional Lie algebra $F_4$ in \cite{O},
acts on the quantum equation via B\"acklund transformations: solutions
to the equation can be mapped to solutions to a transformed equation
obtained by an action of $W(F_4^{(1)})$ on
the weights and the parameter $x$. 
By our discussion above, these results
of \cite{Nagoya:2012tv} predict the transformation properties
of the gauge theory instanton partition function in the presence of a
surface operator under the action of this symmetry group.

We can compare the symmetry group $W(F_4^{(1)})$ of quantum
Painlev\'e, and hence of the five-point function, to the symmetry
$W(D_4) \rtimes SL(2,\IZ)$ of the $\epsilon$-undeformed gauge theory
reviewed in section (\ref{rev_nf4}). The affine Weyl group of $F_4$
can be expressed as $W(F_4^{(1)}) = W(D_4^{(1)}) \rtimes S_4$, with
$\mbox{Aut}_D (D_4^{(1)})=S_4$ the group of automorphisms of the affine Dynkin diagram of $D_4^{(1)}$ \cite{boalch}.
As the $\Gamma(2)$ factor in $SL(2,\IZ) = \Gamma(2) \rtimes
S_3$ is invisible when the theory is expressed in terms of the cross ratio $x$, 
we recognize that the two symmetry groups differ by two
generators, due to replacing the algebra $D_4$ by its affine counterpart $D_4^{(1)}$, and the group of
Dynkin diagram automorphisms $\mbox{Aut}_D(D_4) =S_3$ by its affine counterpart
$\mbox{Aut}_D(D_4^{(1)})=S_4$. Considering the action of the symmetry
generators on the weights and the variable $x$, we observe that the
Weyl symmetry exchanging $m_2$ and $m_3$ of the gauge theory is
deformed to the following symmetry
\begin{eqnarray}  \label{deformed_transf}
(m_1,m_2,m_3,m_4) & \rightarrow & (m_1,m_3 - \frac{\epsilon_2}{2},m_2 + \frac{\epsilon_2}{2},
m_4)
\end{eqnarray}
of the five-point correlator. This deformation is also responsible for the occurrence of the two additional generators. In the $\epsilon_2 \rightarrow 0$ limit, they cease to be independent. Note that the {\it four} point correlator retains the undeformed symmetry, as exemplified in section \ref{ell_rec_rel}.

The transformation properties of the partition function under $W(F_4^{(1)})$ mainly involve rescaling. The transformation (\ref{deformed_transf}) however induces Laplace transformations with regard to the insertion point of the degenerate operator, giving rise to a non-local transformation rule for the partition function. It would be very interesting to verify this transformation directly from the gauge theory perspective.

\subsection{The null vector decoupling equation in elliptic form}
\label{decouplingontorus}
The four-point conformal block depends on the cross ratio $x$ of the
four insertion points. As discussed in section \ref{intro_4pt}, the
four insertion points can be mapped to the 2-torsion points (including
the trivial one) of a torus, such that the modular parameter of the
torus is related to the cross ratio $x$ via relation (\ref{x_of_tau}).

To study the null vector decoupling equation (\ref{null_vect_one}) of
the five-point correlator, it proves advantageous to map this equation
to elliptic variables, extending the map between the 2-torsion points
on the torus and the insertion point of the four-point function to a
two-to-one map between torus and sphere.  This map was performed in
\cite{P}\cite{Zamo1986}\cite{Zamo1987bis} and the result was identified
\cite{Manin}\cite{LO} as an elliptic integrable system
of Inozemtsev type $BC_1$ \cite{Olshanetsky:1981dk}\cite{Inozemtsev}. 
The latter in turn is
equivalent to a Gaudin system \cite{Gaudin} with four punctures after a reduction by an
involution \cite{Zotov} (a
relation that goes back to \cite{F}\cite{S}).
For a history of the derivation, see e.g.
\cite{Takasaki:2000zd}. The reference
\cite{Zabrodin:2011fk} provides further details for the long
calculation, and the reference \cite{Fateev:2009me} states the full
result in the conformal field theory context (for the heavy second
order degenerate insertion).  We provide many technical details of the
derivation in our conventions and starting from the conformal field
theory null vector decoupling equation in appendix
\ref{ellipticvariables}. Here we summarize the main features of the
derivation.

We introduce the coordinate $u$ on the torus via
\be \label{def_z}
z = \frac{\wp(u)-e_3}{e_1-e_3} \,.
\ee

The ansatz
\be  \label{simp_ansatz}
\Psi_5(z,x) = 
\frac{\left[ (1-z) z  (z-x)\right]^{\frac{1}{4}+\frac{b^2}{2}} 
}{ \left[4 x(1-x) \right]^{\frac{1}{12}(1+3b^2+8 \hc)}}  \frac{\vartheta_1(u)^{b^2}}{\vartheta_1'(0)^{\frac{1}{3}(1+b^2)}} 
\prod_{i=1}^3  \left(\frac{  \eta  
}{\vartheta_{i+1}(0)} \right)^{4 h_{i+2}}\psi(u,\tau) 
\ee
for the five-point correlator  reduces the null vector decoupling
equation (\ref{null_vect_one}) to a Schr\"odinger equation for
$\psi(u,\tau)$,
\be  \label{null_vect_ell}
\left(\partial_u^2 + 4 b^2 \sum_{i=0}^3 \hat{h}_i \wp(u+ \omega_i) \right) \psi(u,\tau) =  - 4 \pi i b^2 \partial_\tau  \psi (u,\tau)    \,.
\ee
Here, and in the following, we identify indices ${}_i$ mod $4$. 
The symbols
$\omega_i$ label the 2-torsion points of the torus (with
$\omega_0=0$ the trivial one). The parameters $\hat{h}_i$ are related
to the conformal dimensions $h_i$ via a  shift,
\begin{eqnarray}
\hat{h}_i &=& h_i - \frac{b^2}{4} - \frac{3}{16 b^2} - \frac{1}{2} \,.
\end{eqnarray}
We refer to appendix \ref{ellipticvariables} for the detailed
derivation. 
We note that the weight dependent part of the prefactor in equation (\ref{simp_ansatz}),
\be
\left[4 x(1-x)\right]^{-\frac{2}{3}h_2} \prod_{i=1}^3 \left(\frac{  \eta }{\vartheta_{i+1}(0)} \right)^{4 h_{i+2}} = 
2^{- \frac{4}{3} \sum_{i=0}^3 h_i }
[x(1-x)]^{\frac{h_0}{3}} \left[\frac{x}{(1-x)^2}\right]^{\frac{h_1}{3}} \left[x(1-x)\right]^{-\frac{2}{3}h_2}  \left[\frac{1-x}{x^2} \right]^{\frac{h_3}{3}} \,,
\nonumber 
\ee
transforms as the four-point correlator under channel duality.

\subsection{The monodromy}
We will be searching for very particular solutions to the elliptic
differential equation.  To solve for the four-point conformal block,
we need to impose that the momentum exchanged between the insertions
at $0$ and $x$ has a given
 value $a$, before the exchanged
excitation bifurcates again into the states inserted at $1$ and
$\infty$. To avoid excess notation, we will indicate that this
projection on to intermediate momentum $a$ has taken place simply by
including the argument in the notation. Hence $\Psi_4$ will indicate
the correlator, $\Psi_4(a)$ the corresponding conformal block.

We consider the operator product expansion of $V_{\hd}(0)$ and
$V_{\hc}(x)$, yielding operators $V_\alpha(x)$, and then the operator product expansion of
the degenerate insertion with each of these operators:  
\ba
V_{h_{21}}(z) V_{\alpha}(x) & =& C_{21 , h_\alpha}^{h_{\alpha_+}} (z-x)^{h_{\alpha_+} - h_{21} - h_\alpha} V_{\alpha_+}(x) + \ldots  \label{deg_OPE}\\
&&+ \,C_{21 , h_\alpha}^{h_{\alpha_-}} (z-x)^{h_{\alpha_-} - h_{21} - h_\alpha} V_{\alpha_-}(x) + \ldots \,,   \nn
\ea 
with $\alpha_{\pm} = \alpha \pm \frac{b}{2}$. The monodromy which arises
 when the degenerate insertion at $z$ circles the
operators at $x$ and $0$ is determined by the exponent
\be h_{\alpha_\pm} -h_{21} - h_\alpha
= \frac{b^2}{2} \pm \frac{ab}{\sqrt{\epsilon_1 \epsilon_2}} +
\frac{1}{2} \,,
 \ee
where we have set
$h_\alpha= \frac{Q^2}{4} - \frac{a^2}{\epsilon_1  \epsilon_2}$. 

In terms of the elliptic variable $u$ introduced in (\ref{def_z}) of
periods $\omega,\omega'$, encircling the points $0$ and $x$ on the
sphere in anti-clockwise direction maps to traversing the cycle $[
\kappa \,\omega', \kappa\, \omega' + \omega]$ with $0 < \kappa <
\frac{1}{2}$. The prefactor in (\ref{simp_ansatz}) relating $\Psi_5$
to $\psi$ induces a monodromy $\exp[\pi i(b^2+1)]$.
The monodromy we must impose on the function $\psi(u,\tau)$ is hence
\be \label{monod}
\exp \left[   \pm \frac{2 \pi i ab}{\sqrt{\epsilon_1 \epsilon_2}}  \right] \,.
\ee

Note that a five-point conformal block should ordinarily depend on two intermediate weights. When one insertion is degenerate, the constrained operator product expansion (\ref{deg_OPE}) replaces one such dependence by a discrete label, reflected in the choice of sign
in (\ref{monod}).

\section{The genus expansion}  \label{gen_exp}
To make contact with the topological string partition function in the field theory limit and the quasi-modular behavior of the conformal block that it predicts, we would like to compute the four-point spherical block in an expansion in the topological string coupling $g_s$ and the deformation parameter $s$, related to the $\epsilon$-parameters via 
\be g_s^2 = \epsilon_1 \epsilon_2 \,, \quad s =
(\epsilon_1+\epsilon_2)^2 \,.  \ee 
After reviewing the structure of quasi-modular forms for the group
$\Gamma(2)$ of interest, we will discuss the appropriate limit in the
conformal field theory to implement the genus expansion. We
will then review a recursion relation \cite{Zamo1987bis} which permits
the computation of the conformal block in this limit, thus yielding the corresponding topological string partition function,
order by order in $q_2$
 and to all orders in the parameters $g_s$ and $s$.
By expanding in $g_s$ and $s$ and imposing quasi-modularity, we can determine all order results in $q_2$ after a finite number of recursion steps. As a proof of principle, we compute $F^{(n,g)}$, $n+g=2$, in this manner.

In section \ref{semi_class_null_vect},
 we turn
to the solution of the null vector decoupling equation
(\ref{null_vect_ell}) in the genus expansion limit.
 Extracting the four-point block
from this computation requires taking the strict $\epsilon_2
\rightarrow 0$ limit, yielding modular results order by order in the
deformation parameter $s$ and at leading order in the topological
string coupling $g_s$, and to all orders in the modular parameter $q_2$.

\subsection{Quasi-modular forms of $\Gamma(2)$} \label{quasi} The
invariance of the $N_f=4$ theory would naively suggest that the
$\tau$-dependence of all observables holomorphic in $\tau$ should be
captured in terms of modular forms of $\Gamma(2)$. These form the
space $M(\Gamma(2))$, which is spanned by the powers of
theta-functions $\vartheta_2^4$ and $\vartheta_3^4$ of weight 2, or
equivalently, by the half-periods $e_1$ and $e_2$.  Via the relation
to topological strings, we know that quantities naively holomorphic in
$\tau$ may in fact exhibit $\bar{\tau}$ dependence. A minimal extension of
$M(\Gamma(2))$ to allow for this is the space of almost holomorphic
modular forms $\widehat{M}(\Gamma(2))$. The elements of this space are
polynomials in $1/ \im(\tau)$, with coefficients holomorphic in
$\tau$, which transform modularly under $\Gamma(2)$ (see
e.g. \cite{KZ}). The space $\widehat{M}(\Gamma(2))$ is obtained from
$M(\Gamma(2))$ by adding the non-holomorphic Eisenstein series $\hat{E_2}$
as a further generator. A quasi-modular form of weight $k$ is the constant term
in the $1/ \im(\tau)$ expansion of an almost holomorphic modular form
of weight $k$. The rings of almost holomorphic modular forms
$\widehat{M}(\Gamma(2))$ and of quasi-modular forms
$\widetilde{M}(\Gamma(2))$ are isomorphic, the isomorphism given by
mapping the generator $\hat{E_2}$ to the holomorphic (non-modular)
Eisenstein series $E_2$. The space of quasi-modular forms
$\widetilde{M}(\Gamma(2))$ is closed under differentiation.
Quasi-modular forms of $\Gamma(2)$ are naturally expanded in terms of the modular variable $q_2$, 
where $q_n = \exp \frac{2 \pi i \tau}{n}$.

By imposing quasi-modularity, we will be able to extract all
order results in $q_2$ from the elliptic recursion relation reviewed in section
\ref{ell_rec_rel}. In our computations of the four-point
spherical block in section \ref{semi_class_null_vect},
quasi-modularity will be manifest for the derivative of coefficients of the logarithm of
the semi-classical conformal block (via evaluation of period integrals
of the Weierstrass function $\wp$) in a genus expansion. We will then verify experimentally that these results also integrate to quasi-modular forms.

\subsection{The genus expansion limit} \label{genus_limit}
The refined topological string partition function is defined as the following generating function of the topological string amplitudes:
\be \label{Z_top}
Z_{top} = \exp \sum_{n,g=0}^\infty F^{(n,g)} s^n g_s^{2g-2}  \,.
\ee  
It is natural to consider the $\epsilon_1, \epsilon_2 \rightarrow 0$
limit in the conformal field theory to reproduce this expansion. This amounts
 via the 2d/4d dictionary (\ref{dictionary})
to the condition that the exchanged
conformal dimension $h$ be much bigger than the central charge $c$,  $h
\gg c$. The leading
behavior of the partition function (\ref{Z_top}) in this limit is reminiscent of the
semi-classical limit $b \rightarrow 0$ of the conformal block. We
 know, through an analysis of the null vector decoupling equation,
that the leading term in the semi-classical conformal block is of
order $\exp (a^2/\epsilon_1 \epsilon_2)$ in this limit
\cite{Zamo1986}. Thus, we expect a semi-classical four-point spherical
block of the form
\be 
\label{genus_exp} 
\Psi_4(a) = \exp
\left[\frac{1}{\epsilon_1 \epsilon_2} F(\epsilon_1,\epsilon_2) \right]
\,, 
\ee 
with $F$ a formal power series in $\epsilon_1$ and $\epsilon_2$.  Note
that equating this block with (\ref{Z_top}) predicts constraints on
the terms in the power series $F$. Aside from the symmetry between
$\epsilon_1$ and $\epsilon_2$ that is naively broken by the
semi-classical limit, the absence of half-integer powers of the
deformation parameter $s$ is non-trivial. Indeed, to reproduce this
integral expansion from the $\epsilon$-deformed partition function
of \cite{Nekrasov:2002qd}
requires an $\epsilon$-dependent shift of the mass parameters
of \cite{Nekrasov:2002qd}. The integral expansion arises
naturally in the context of the holomorphic anomaly equations of
topological string theory \cite{Huang:2011qx}, and can be reproduced
in conformal field theory both from the elliptic recursion relations
and the null vector decoupling equations, as we discuss in this
section.

\subsection{The elliptic recursion relation} 
\label{ell_rec_rel}
Conformal blocks are functions of the central charge $c$ of the
theory, as well as of the exchanged weight $h$. Considered as
functions of the central charge $c$ for fixed $h$, they exhibit poles
at those values $c_{mn}$ of the central charge $c$ at which $h$
corresponds to a degenerate weight of degree $mn$. Considered as
functions of the exchanged dimension $h$ for fixed $c$, they exhibit
poles at degenerate weights $h_{mn}(c)$. In both cases \cite{Zamolodchikov:1990ww}, the residue of
the poles is proportional to the conformal block itself, evaluated at
the respective central charge or singular weight. 

For the four-point conformal block on the sphere, making the poles in
the central charge $c$ explicit gives rise to the formula
\be 
F(c,h,h_i,x) = f(h,h_i,x) + \sum_{m,n}
\frac{R'_{mn}(h,h_i)}{c-c_{mn}(h)} F(c_{mn},h+mn,h_i,x)
\,, \label{recursion_x} 
\ee 
where the function $R'_{mn}(h,h_i)$ grows polynomially
with $h$. The convergence of this expansion hence requires $c \gg
h$. The function $f(h,h_i,x)$ is the infinite central charge limit of the conformal
block. It can be determined in terms of hypergeometric functions to be
\be f(h,h_i,x) = x^{h-\hc-\hd} {}_2 F_1(h+\hc-\hd, h+ \ha-\hb, 2h, x)
\,.  \ee Due to the factor $x^h$ in the leading behavior
 and the shift $h
\rightarrow h+ mn$ on the right-hand side of (\ref{recursion_x}), this equation 
gives rise to a recursion relation to determine the conformal block
order by order in the cross ratio $x$ \cite{Zamolodchikov:1985ie}.
 By the 2d/4d dictionary, this yields the
instanton expansion on the gauge theory side of the correspondence. In
particular, the $\epsilon$-deformed partition function \cite{Nekrasov:2002qd} yields a closed form
solution to this recursion relation.

The genus expansion emerges from the second point of view. It proves convenient to
write the conformal block as a product of functions of the elliptic
parameter $q_2$,
\be \label{factor_block_rec} F(c,h,h_i,q_2)
= f(c,h,h_i,q_2) H_h(c,h_i,q_2) \,.  
\ee 
The rationale behind replacing
the cross ratio $x$ dependence by $q_2$ dependence, with $q_2=e^{\pi i \tau}$ and $\tau$
and $x$ related via (\ref{x_of_tau}), will become clear
momentarily. 

The first factor $f(c,h,h_i,q_2)$ yields the leading behavior of
the conformal block in the $h \gg c \gg 1$ limit. It is computed
in \cite{Zamo1986}\cite{Zamo1987bis} and given by 
\ba
f(c,h,h_i,q_2) &=& (16 q_2)^{h - \frac{c-1}{24}} x^{ \frac{c-1}{24}- \hc - \hd} (1-x)^{ \frac{c-1}{24}- \hc - \hb} \vartheta_3(0)^{  \frac{c-1}{2}- 4 ( \ha+\hb+\hc+\hd)} \\
&=& (16 q_2)^{- \frac{a^2}{\epsilon_1 \epsilon_2}}
\left(\frac{\vartheta_3^2}{\vartheta_0 \vartheta_2 \vartheta_3}
\right)^{4h_2} \left( \vartheta_0 \vartheta_2 \vartheta_3
\right)^{Q^2} \prod_{i=1}^3 (\vartheta_{i+1})^{-4 h_{i+2}}
\,. \label{Z_leading_q} 
\ea 
 The second factor in
the block (\ref{factor_block_rec}) satisfies the relation
 \be
 \label{recursion_q} H_h (c,h_i, q_2) = 1 + \sum_{m,n} (16 q_2)^{mn} \frac{R_{mn}(c,h_i)}{h -
  h_{mn}(c)} H_{h_{mn}+mn}(c,h_i,q_2) \,, 
\ee 
with  the residue factor given by $R_{mn}(c,h_i) = A_{mn}
\prod_{i=1}^4 Y_{rs}(2m_i)$, and the functions $A$ and $Y$ by
 \be \label{residue} A_{mn} = \frac{1}{2}
\left( \prod_{r=1-m}^m \prod_{s=1-n}^n \right)' \frac{1}{r b + s
  b^{-1}} \,,\quad
Y_{mn}(M) =  \left(\prod_{r=1-m}^{m-1} \prod_{s=1-n}^{n-1} \right)''   (\frac{M}{\sqrt{\epsilon_1 \epsilon_2}} -\frac{r b + s b^{-1}}{2})  \, .
\ee 
The prime on the first product indicates that the factors $(r,s) =
(0,0),(m,n)$ are to be omitted, and the double prime on the second
product prescribes a product over pairs satisfying $ (r,s)= (1-m,1-n)
\, \mbox{mod} \, (2,2)$. The conformal dimensions $h_{mn}$ of
degenerate representations are given in equation
(\ref{deg_weights}). The leading behavior $q_2^h$ of the
conformal block \cite{Zamo1986} gives rise to a recursion relation with regard to the
order in the elliptic parameter $q_2$, providing the
 rationale for introducing the elliptic parameter \cite{Zamo1987bis}.

We will now exploit the elliptic recursion relation in various ways.
Firstly, we note that the existence of the recursion relation allows
us to rule out the convergence of the genus expansion
(\ref{Z_top}) of the conformal block. This is most simply stated in
the massless limit. In this limit, the partition function
(\ref{Z_top}) takes the form 
\be \label{double_sum} \log
Z_{top} = \frac{1}{\epsilon_1 \epsilon_2} \sum_{r,s} \left( \sum_n
  \psi_{n}^{rs} q_2^n \right) \epsilon_1^r \epsilon_2^s \,.
\ee 
The coefficients $F^{(n,g)}(q_2) = \sum_n \psi_{n}^{rs}
\,q_2^n$
sum to quasi-modular forms. The sum over $q_2$ is hence convergent in the upper
half-plane. If the double sum (\ref{double_sum}) were convergent, hence absolutely convergent within the radius of convergence,
 we would be justified in reversing the order of
summation. This is the order in which the recursion relation yields
the conformal block, with 
\be \label{epsilon_summed} \sum_{rs}
\psi_n^{rs} \,\epsilon_1^r \epsilon_2^s =
\frac{p_{n}(\epsilon_1,\epsilon_2)}{q_{n}(\epsilon_1,\epsilon_2)}
\ee 
a rational function in the $\epsilon$-parameters. As one easily
verifies from the explicit expressions (\ref{residue}), the
$\epsilon$-parameters in the function $q_{n}(\epsilon_1,\epsilon_2)$ are
multiplied by integers of magnitude increasing with $n$. It
follows that the radius of convergence of the series in $\epsilon_1$
and $\epsilon_2$ on the left-hand side of equation (\ref{epsilon_summed})
decreases as $n$ grows. For any choice of the
$\epsilon$-parameters, it ceases to converge for sufficiently large $n$. We conclude that the double sum (\ref{double_sum}) cannot be
convergent. 

The restriction to the massless case was not
essential to this argument. In the massive case, the sum in
parentheses in (\ref{double_sum}) is replaced by a more intricate
expression. Via the holomorphic anomaly equation, it can be expressed in closed
form as a quasi-modular form in a modular variable different from $q_2$
(namely the complex structure of the corresponding massive Seiberg-Witten
curve), with coefficients depending on all parameters of the
theory. Alternatively, it can be expressed as infinite sums over the
mass parameters, with coefficients quasi-modular in $q_2$.
 Either way,
the argument yielding non-convergence still holds.

As a second application of the elliptic recursion relation (\ref{recursion_q}),
we note that it yields a purely conformal field theoretic method of determining
the topological string amplitudes $F^{(n,g)}$ in a mass expansion,
with coefficients that are quasi-modular forms of $\Gamma(2)$, the
subgroup of the S-duality group that does not act on the masses. At
order $m$ in the masses, the weight of the quasi-modular coefficient
is $w=2(n+g-1)+m$. As the space of quasi-modular forms of this weight
is a vector space of dimension $\binom{\frac{w}{2}+2}{2}$, the exact
expression for the partition sum $F^{(n,g)}$ at a given order in the masses can be
determined from this number of coefficients in $q_2$, hence after a
finite number of recursion steps. It follows from (\ref{residue}) that
the maximal order in the mass parameters is bounded at any given order
in $q_2$. Hence, the quasi-modular coefficients of terms of high order
in mass must conspire such that low orders in the modular parameter
cancel. Analogous considerations apply to the case of the toroidal
one-point block discussed in \cite{KashaniPoor:2012wb}.

To demonstrate the utility of this approach in practice, we determined 
the partition sums $F^{(n,g)}$ for $n+g=2$, to order $a^{-6}$ in the exchanged
conformal dimension from the elliptic recursion relation. The results are
\begin{eqnarray}
F^{(0,2)} &=& \frac{1}{32 a^2} E_2 + \frac{1}{2880 a^4}
(95 E_2^2 + 49 E_4)R + \frac{1}{a^6} 
( ( \frac{145}{5184} E_2^3 
+ \frac{169}{4320}  E_2 E_4
+ \frac{421}{     25920}  E_6) R^2 
\nonumber \\
& & 
+ (-\frac{11}{64 \pi^2} E_2^2 
- \frac{31}{
     192 \pi^2} E_4) (e_1 T_1 + e_2 T_2 + e_3 T_3) 
- \frac{1}{2 \pi^4} E_2
   (e_1^2 T_1 + e_2^2 T_2 + e_3^2 T_3)) \,,
\nonumber \\
F^{(1,1)} &=&
- \frac{1}{24 a^2} E_2 - \frac{1}{360 a^4} (10 E_2^2+ 11 E_4) R
+ \frac{1}{a^6} ((-\frac{1}{54} E_2^3 
- \frac{11}{240}  E_2 E_4
- \frac{71}{2160}  E_6) R^2
\nonumber \\
& & 
+ (\frac{11}{
     144 \pi^2} E_2^2
+ \frac{5}{16 \pi^2} E_4) (e_1 T_1 + e_2 T_2 + e_3 T_3) 
+  \frac{7}{12 \pi^4} E_2 (e_1^2 T_1 + e_2^2 T_2 + e_3^2 T_3)) \,,
\nonumber \\
F^{(2,0)} &=&
 \frac{1}{96 a^2} E_2+ \frac{1}{960 a^4} ( 5 E_2^2 + 9 E_4) R
+ \frac{1}{a^6} ((
\frac{5}{1728} E_2^3
+ \frac{7}{720}  E_2 E_4
+ \frac{101}{     8640} E_6) R^2
\nonumber \\
& & 
+ (-\frac{5}{576} \pi^2 E_2^2
- \frac{7}{
     64 \pi^2}  E_4)(e_1 T_1 + e_2 T_2 + e_3 T_3) 
- \frac{1}{12 \pi^4}
  E_2 (e_1^2 T_1 + e_2^2 T_2 + e_3^2 T_3)) \,.
\nonumber 
\end{eqnarray}
Up to order $a^{-4}$, these results match the results of
\cite{Billo:2013fi} determined using the recursion relation in the
cross ratio $x$. The term in $F^{(2,0)}$ of order $a^{-6}$ matches the
result we will determine using the null vector decoupling equation.

\subsection{Solving the null vector decoupling equation in the genus expansion} \label{semi_class_null_vect}

To solve the null vector decoupling equation (\ref{null_vect_ell}) in the semi-classical limit,
we use the following ansatz motivated by Liouville theory (as explained e.g. in \cite{KashaniPoor:2012wb}):
\begin{eqnarray} \label{ansatz_psi}
\psi(u,\tau) &=& \exp \left[ \frac{1}{\epsilon_1 \epsilon_2} 
{\cal F}(\tau) + \frac{1}{\epsilon_1} {\cal W}(u,\tau)  + O(\epsilon_2) \right] \,.
\end{eqnarray}
With the dictionary (\ref{dictionary}), the differential equation in the $\epsilon_2 \rightarrow 0$ limit becomes
\be 
-\frac{1}{\epsilon_1}
\cW''(z|\tau) - \frac{1}{\epsilon_1^2} \cW'(z|\tau)^2 + \sum_{i=0}^3 \left(
  \frac{4}{\epsilon_1^2} M_i^2 -\frac{1}{4} \right) \wp(z+ \omega_i) = 2 (2 \pi i)^2
\frac{1}{\epsilon_1^2} q \partial_q {\cal F} (\tau)
\,.  \label{eq_in_w} 
\ee 
Note that only the derivative of ${\cal F}$ appears, as the normalization of $\psi(u,\tau)$ is neither fixed by the differential equation nor by the monodromy constraint. 

Making a power series ansatz in $\epsilon_1$ for both $\cF(\tau)$ and $\cW(z|\tau)$,
\be
 {\cal F}(\tau) =
\sum_{n=0}^\infty {\cal F}_n(\tau) \epsilon_1^n \,, 
\quad {\cal  W}(z|\tau) = \sum_{n=0}^\infty {\cal W}_n(z|\tau) \epsilon_1^n \,,   \label{pert_ansatz}
\ee 
we obtain the system of equations
\beq
-{\cW'_0}^2 + 4 \sum_{i=0}^3 M_i^2 \wp(z + \omega_i)  &=&2 (2 \pi i)^2 q \partial_q \cF_0 \,, \label{first_eq} \\
-\cW''_0 - 2 \cW'_0 \cW'_1&=& 2(2 \pi i)^2 q \partial_q \cF_1 \,, \\
- \cW''_1 - {\cW'_1}^2  -2 \cW'_0 \cW'_2 - \frac{1}{4} \sum_{i=0}^3 \wp(z+ \omega_i) &=& 2 (2 \pi i)^2 q \partial_q \cF_2 \,,\label{third_eq}\\
- \cW''_{n} - \sum_{i=0}^{n+1} \cW'_i \cW'_{n+1-i} &=& 2(2 \pi 
i)^2 q \partial_q \cF_{n+1}  \quad \mbox{for} \quad {n \ge 2} \,.
\label{generic_eq}
\eeq 
The monodromy condition is imposed by demanding
\be
\oint \cW_0' = \pm 2 \pi i a \,, \quad \oint \cW_i' = 0  \quad \mbox{for} \quad i>0
\,.  \label{bc} \ee

The solution of the system of equations (\ref{first_eq}) --
(\ref{generic_eq}) proceeds much as in \cite{KashaniPoor:2012wb}. For
the massless case, we can solve exactly in terms of quasi-modular
forms in $\tau$. (Exceptions to non-quasi-modularity are the terms in $\cF_0$ and $\cF_2$ 
at order $a^2$ and $a^0$.) For the massive case, we
define a second perturbation expansion in the parameters
\begin{eqnarray}
v_i &=& \left(\frac{M_i}{\pi a} \right)^2 \, .
\end{eqnarray}
The contour integrals that one needs to perform to impose the monodromy conditions
(\ref{bc}) generalize those of \cite{KashaniPoor:2012wb}.
In particular, the integrands include products of the Weierstrass $\wp$-function and their derivatives evaluated at
arguments shifted by 2-torsion points.
We derive a 
 set of recursion
relations to calculate the monodromies of such integrands in appendix \ref{Weierstrassmonodromies}.
We can thus compute $\cF(\tau)$, up to an integration constant, recursively order by order in $\epsilon_1$ and the $v_i$, and exactly at each such order in the modular parameter $q_2$.

Our solution algorithm hence
 gives rise to $q \partial_q \cF_n$ as
elements of the ring 
\be \label{ring} q \partial_q \cF_n \in
\IC\left[\frac{1}{a}\right][E_2, e_1,
e_2]\bigg[\bigg[\frac{M_i}{a}\bigg]\bigg] \,, \quad n>2 \,.  
\ee
In fact, more is true. As predicted by the 2d/4d correspondence and
analogously to the observations in the case of the torus one-point
function in \cite{KashaniPoor:2012wb}, $q \partial_q \cF_n$ lies in the subring of derivatives of quasi-modular forms, such that
$\cF_n$ itself is an element of the right-hand side of
(\ref{ring}). We have to date been able to verify this latter property
merely experimentally. 

The statement that $q \partial_q \cF_n$ integrates to a quasi-modular
form is of course only true for an appropriate choice of integration
constant. For ease of notation, we shall cite our results below with
this choice implemented, which we indicate by a superscript `c' (for
`complete'). Given the normalization (\ref{intro_conf_block}), the
proper choice of integration constant however is zero; the constant
terms of the quasi-modular forms are provided by the product
of Liouville three-point functions multiplying the conformal blocks in
equation (\ref{intro_conf_block}).

As anticipated in section \ref{quasi}, the
subgroup $\Gamma(2)$ of the S-duality group that acts only on $\tau$
is realized by the fact that the $\tau$-dependence of our
results is in terms of quasi-modular forms of
$\widetilde{M}(\Gamma(2))$. The full S-duality group $SL(2,\IZ)
\rtimes W(D_4)$
 is realized in an intricate manner already at play in
\cite{Seiberg:1994aj}: the action of $\Gamma(2)$ on $e_1$ and $e_2$
can be extended to that of $SL(2,\IZ) = \Gamma(2) \rtimes S_3$ on
$(e_1,e_2,e_3=-e_1-e_2)$, with $S_3$ permuting the entries of this
vector.  Expressing our results in terms of the $SO(8)$ Casimirs (\ref{casimirs})
 makes the full S-duality symmetry manifest.

The proof that the odd terms vanish, $q \partial_q \cF_{2n+1} = 0$, proceeds as in
\cite{KashaniPoor:2012wb}.
 The non-vanishing results we obtain, to
order six in $\epsilon_1$, are the following:\footnote{We thank the authors of \cite{Billo:2013fi} for pointing out some typos in the following expressions in a previous version of this paper.}
\begin{align}
{\cal F}_0^c &= - a^2 \log q_2 + 4 R \log \eta +
\frac{1}{a^2} \left(\frac{E_2 R^2}{6}-\frac{e_1 T_1+e_2 T_2+e_3 T_3}{\pi ^2} \right) \nn \\
&+\frac{1}{a^4} \bigg[\frac{E_4 N}{5}+\left(\frac{E_2^2 }{36}+\frac{E_4}{180} \right)R^3 
-\frac{E_2 R \left(e_1 T_1+e_2 T_2+e_3 T_3\right)}{3 \pi ^2}-\frac{R \left(e_1^2 T_1+e_2^2 T_2+e_3^2 T_3\right)}{2 \pi ^4}\bigg] \nn\\
&+\frac{1}{a^6}\bigg[\frac{2}{15}
E_2 E_4 N R+\frac{2 E_6 N R}{35}+\frac{5 E_2^3 R^4}{648}+\frac{1}{270} E_2 E_4 R^4+\frac{11 E_6 R^4}{22680} \nn\\
&-\frac{5
E_2^2 R^2 \left(e_1 T_1+e_2 T_2+e_3 T_3\right)}{36 \pi ^2} -\frac{E_4 R^2 \left(e_1 T_1+e_2 T_2+e_3 T_3\right)}{12 \pi ^2}+\frac{E_2
\left(e_1 T_1+e_2 T_2+e_3 T_3\right){}^2}{6 \pi ^4} \nn \\
&-\frac{E_2 R^2 \left(e_1^2 T_1+e_2^2 T_2+e_3^2 T_3\right)}{3 \pi ^4}+\frac{4}{189} E_6
\left(T_1 T_2+T_1 T_3+T_2 T_3\right)+\frac{e_1^3 T_1^2+e_2^3 T_2^2+e_3^3 T_3^2}{\pi ^6}\bigg]\nn\\
& + \mbox{higher order terms in the masses,} \nn \\
& \nn \\
{\cal F}_2^c &= - \log \eta -\frac{E_2 R}{12 a^2}+\frac{1}{a^4} \left[ -\frac{1}{48} \left( E_2^2+E_4  \right)R^2 +\frac{E_2 \left(e_1 T_1+e_2 T_2+e_3 T_3\right)}{12
\pi ^2}+\frac{e_1^2 T_1+e_2^2 T_2+e_3^2 T_3}{2 \pi ^4} \right] \nn \\
&+\frac{1}{a^6} \bigg[-\left( \frac{1}{30} E_2 E_4 +\frac{2 E_6 }{15}\right)N - \left(\frac{5 E_2^3
}{648}+\frac{2}{135} E_2 E_4 +\frac{17 E_6 }{3240} \right)R^3 \nn\\
&+ \left( \frac{5 E_2^2 }{72 \pi
^2}+\frac{5 E_4 }{24 \pi ^2} \right) R \left(e_1 T_1+e_2 T_2+e_3 T_3\right)+\frac{5 E_2 R \left(e_1^2 T_1+e_2^2 T_2+e_3^2 T_3\right)}{12 \pi ^4} \bigg]  \nn \\
&+ \mbox{higher order terms in the masses} \,, \nn \\
& \nn \\
{\cal F}_4^c &= \frac{E_2}{96 a^2}+\frac{1}{a^4} \left(\frac{E_2^2 }{192}+\frac{3 E_4 }{320} \right)R+\frac{1}{a^6} \bigg[ \left(\frac{5 E_2^3 }{1728}+\frac{7}{720}
E_2 E_4 +\frac{101 E_6}{8640} \right) R^2 \nn\\
& - \left(\frac{5 E_2^2 }{576 \pi ^2}+\frac{7 E_4 }{64 \pi ^2} \right) \left(e_1 T_1+e_2 T_2+e_3 T_3\right) -\frac{E_2}{12 \pi ^4}\left(e_1^2 T_1+e_2^2 T_2+e_3^2 T_3\right)\bigg] \nn\\
&+\frac{1}{a^8} \bigg[\left(\frac{7 E_2^2
E_4 }{1440}+\frac{239 E_4^2 }{1440}+\frac{E_2 E_6 }{30} \right) N  +  \left( \frac{35 E_2^4 }{20736}+\frac{203 E_2^2 E_4
}{25920}+\frac{1193 E_4^2 }{103680}+\frac{337 E_2 E_6 }{25920} \right) R^3 \nn \\
&  - \left(\frac{7 E_2^3 R }{576
\pi ^2}+\frac{161 E_2 E_4 R }{960 \pi ^2}+\frac{77 E_6 R }{480
\pi ^2} \right) \left(e_1 T_1+e_2 T_2+e_3 T_3\right)\nn \\
&- \left(\frac{7 E_2^2 }{64 \pi ^4}+\frac{77 E_4}{192
\pi ^4}\right)  \bigg] R \left(e_1^2 T_1+e_2^2 T_2+e_3^2 T_3\right)  + \mbox{higher order terms in the masses} \,, \nn \\
& \nn \\
{\cal F}_6^c &=-\frac{1}{a^4}\left(\frac{E_2^2}{2304}+\frac{13 E_4}{11520}\right)
-\frac{1}{a^6}\left(\frac{5 E_2^3 }{10368}+\frac{E_2 E_4 }{432}+\frac{355
E_6 }{72576}\right)R \nn \\
&+\frac{1}{a^8} 
\bigg[ \left(-\frac{35 E_2^4 }{82944}-\frac{287 E_2^2 E_4 }{103680}-\frac{9235  E_4^2}{580608}-\frac{5671 E_2 E_6 }{725760} \right) R^2 \nn \\ 
&+\left( \frac{7 E_2^3 }{6912 \pi ^2}+  \frac{971 E_2 E_4 }{34560 \pi ^2}+\frac{1267 E_6 }{17280 \pi ^2}\right)\left(T_1 e_1 +T_2 e_2 +T_3 e_3 \right) 
\nn \\
& + \left(\frac{7 E_2^2 }{576 \pi ^4}  + \frac{95 E_4 }{576 \pi ^4}\right) \left(T_1 e_1^2 +T_2 e_2^2 +T_3 e_3^2 \right) \bigg]  \nn  \\
&  +\frac{1}{a^{10}} \bigg[- \left(\frac{ E_2^3
E_4}{1440}+\frac{89  E_2 E_4^2 }{1600}+\frac{1}{160}  E_2^2 E_6 +\frac{4679  E_4 E_6 }{14400} \right) N \nn \\
&- \left(\frac{7
E_2^5 }{20736}+\frac{7 E_2^3 E_4 }{2592}+\frac{22963 E_2 E_4^2 }{907200}+\frac{803 E_2^2 E_6 }{96768}+\frac{28963
R^3 E_4 E_6}{1036800} \right) R^3\nn \\ 
& + \left( \frac{7 E_2^4  }{3456 \pi ^2}+\frac{125  E_2^2 E_4  }{2304 \pi ^2}+\frac{20971 E_4^2 }{48384 \pi ^2}+\frac{1259 E_2 E_6
}{8064 \pi ^2} \right) R\left(T_1 e_1 +T_2 e_2 +T_3 e_3\right)\nn \\
&+ \left( \frac{13 E_2^3 }{576 \pi ^4}+\frac{2077 E_2
E_4 }{5760 \pi ^4} +\frac{3373 E_6 }{5760 \pi
^4} \right) R \left(T_1 e_1^2 +T_2 e_2^2+ T_3 e_3^2 \right) \bigg] 
\nn \\
&  + \mbox{higher order terms in the masses.} \label{null_rec_results}
\end{align}
The derivation of these results 
also required
 determining the quantities ${\cal W}'_n (z)$ to particular orders in $\epsilon_1$ and the mass parameters.

The 2d/4d correspondence implies the identification $F^{(n,0)} =
\cF_{2n}^c$. 
The vanishing of $\cF_{2n+1}$ is hence consistent with the
fact that no half-integer powers of $s$ appear in the expansion
(\ref{Z_top}). Our results in (\ref{null_rec_results}) confirm
the massless results obtained via the holomorphic anomaly relations in
\cite{Huang:2011qx}, as well as the results in \cite{Billo:2011pr} and
\cite{Billo:2013fi}, of which the latter are based on combining instanton calculus
with the demand for quasi-modularity, to the order given
there.\footnote{There are minor typos in the signs in equation (3.22) in
  version 1 of \cite{Billo:2013fi}.} The paper \cite{Huang:2011qx} also gives
closed results for the massive amplitudes, but as a function of both
the UV coupling (the coupling $q_2$
 considered here), and the effective
coupling determined via special geometry. Furthermore, the results are
expressed in terms of the Seiberg-Witten $u$ coordinate, rather than
the flat coordinate $a$. To check against the results of
\cite{Huang:2011qx}, we can generate our power series in masses by
expressing the former purely in terms of $q_2$ and $a$.

The terms in the logarithm of the coefficient of $\psi(u,\tau)$ in
(\ref{simp_ansatz}) that scale as $\frac{1}{\epsilon_2}$ and thus
contribute to $F^{(n,0)}$
are
\be 
-\frac{8}{3}h_2 \log \frac{\vartheta_0 \vartheta_2}{\vartheta_3^2} - 4 \sum_{i=1}^3 h_{i+2} \log \frac{\vartheta_{i+1}}{\eta} = 8 h_2 \log \frac{\vartheta_3}{\eta} - 4 \sum_{i=1}^3 h_{i+2} \log \frac{\vartheta_{i+1}}{\eta} + \mbox{const} \,,
\ee
where the last term is $\tau$-independent. These terms only contribute to $F^{(0,0)}$
and $F^{(1,0)}$, such that
\ba
F^{(0,0)} &=& - 8 M_2^2 \log \frac{\vartheta_3}{\eta}+ 4 \sum_{i=1}^3 M_{i+2}^2 \log \frac{\vartheta_{i+1}}{\eta} + \cF_0^c \\
&=&   - a^2\log q_2  - 4 M_2^2 \log \frac{\vartheta_3^2}{\eta^3} + 4 \sum_{i=1}^3 M_{i+2}^2 \log \vartheta_{i+1} + \mbox{quasi-modular forms}    \,  \nn
\ea
and
\ba
F^{(1,0)} &=& 
 \log 
\frac{\vartheta_3}{\vartheta_0 \vartheta_2} + \mbox{quasi-modular forms}    \,  \nn
\ea
Note that we have thus reproduced the leading behavior (\ref{Z_leading_q}) of the conformal block as determined in \cite{Zamo1986} to leading order in $\epsilon_2$.

\section{Conclusion}
In this paper, we continue our study of the duality between
two-dimensional conformal field theory and four-dimensional ${\cal
  N}=2$ supersymmetric gauge theory \cite{Alday:2009aq} with an
emphasis on the genus expansion. Following
up on our analysis of the one-point toroidal block
\cite{KashaniPoor:2012wb}, we uncover modular properties of
the spherical four-point conformal block.
In particular, we show that when we assume quasi-modularity of the
genus expansion coefficients, the elliptic recursion relations allow
for their straightforward determination. At the same time, the
recursion relation permits us to conclude that the genus expansion is
formal, with vanishing radius of convergence.  We demonstrate
that in the study of the semi-classical limit of an appropriate null
vector decoupling equation, quasi-modularity arises
intrinsically. Upon making a power series ansatz for the five-point
block, we obtain a system of equations for the expansion coefficients which can be studied on its own merits. 
The $\tau$-derivatives of these coefficients are manifestly
quasi-modular. We have experimentally observed that they also
integrate to quasi-modular forms. 

The exact null vector decoupling equation on the five-point function is interesting in its own
right, as it conjecturally governs the instanton
partition function in the presence of an elementary surface
operator \cite{Alday:2009fs}. By exploiting its relation to the quantum Painlev\'e VI
equation \cite{Nagoya:2012tv}, we have been able to determine the transformation properties
of this gauge theory object under the action of the group
$W(D_4^{(1)}) \rtimes \Aut_D(D_4^{(1)})$, an enhancement of the
symmetry group $W(D_4) \rtimes \Aut_D(D_4)$ for the $\epsilon$-undeformed $N_f=4$ gauge theory
\cite{Seiberg:1994aj}.

Our work gives rise to numerous questions worthy of further study. The
predicted transformation properties of the instanton partition
function in the presence of a surface operator need to be understood
from the vantage point of gauge theory. In the same vein, the symmetry
enhancement we observed needs to be investigated in the full
$\epsilon$-deformed gauge theory. Furthermore, the identification of
the null vector decoupling equation with the Inozemtsev integrable
system remains to be fully exploited.
Finally, quasi-modularity is the determining feature underlying our
computations. It should be possible to identify a limit in parameter
space of more general quiver gauge theories in which the introduction
of elliptic or generalized modular variables yields similar
structure. Proving quasi-modularity of the expansion coefficients
determined via the null vector decoupling equation, finding an
interpretation of this symmetry from within conformal field theory,
and developing a better understanding of its occurrence against the
backdrop of the non-convergence of the power series giving rise to it pose important open challenges.

\section*{Acknowledgments}
We would like to thank Yuji Tachikawa for interesting correspondence. Our work is supported in part by the grant ANR-09-BLAN-0157-02. 

\appendix

\section{The null vector decoupling equation in elliptic variables}
\label{ellipticvariables}
In this appendix, we explicitly perform the map between the null
vector decoupling equation in spherical variables to the differential
equation in elliptic variables. Our derivation combines results in
references
\cite{P}\cite{Zamo1987bis}\cite{Manin}\cite{LO}\cite{Takasaki:2000zd}\cite{Fateev:2009me}; the details provided in \cite{Zabrodin:2011fk} proved helpful.

\subsection{The change of variables}
The Weierstrass $\wp$-function defines a two-to-one map from the torus
$\IC/\Lambda$ to the sphere $\IC \cup \infty$, mapping the 2-torsion
points $\omega_i$ to the half-periods $e_i = \wp(\omega_i)$, and the
origin to infinity. As we wish to map these points to the insertion
points $(\infty,1,x,0)$ instead, we consider the map 
\be \label{z_map}
 z = \frac{\wp(u)-e_3}{e_1-e_3} \,, 
\ee 
which in particular maps $\omega_2$
to $x$, 
\be x = \frac{e_3- e_2}{e_3- e_1} (\tau) \,.  \label{map} 
\ee
Note that in the following, $z$ will both specify a general coordinate
on the sphere and the insertion point of the degenerate insertion,
likewise for $u$ and the torus -- we trust that this will not give
rise to confusion. To invert the map (\ref{z_map})
 locally, note that the right-hand
 side is the inverse of the squared Jacobi sine function
introduced in appendix \ref{identities}. We can hence invoke
equation (\ref{inv_sn}) to obtain 
\be 
u = \frac{1}{\sqrt{e_1-e_3}}
\int_0^{\sqrt{\frac{1}{z}}} \frac{dy}{\sqrt{(1-y^2)(1-k^2 y^2)}} \,.
\ee

To convert the null vector decoupling equation
\ba
\Big[-\frac{h_{(2,1)}}{ (z-1) z} + \frac{\ha }{(z-1) z} + \frac{\hb }{(z-1)^2
z} - \frac{\left(x^2+z-2 x z\right) \hc}{(x-z)^2 (z-1) z}- \frac{\hd}{z^2(z-1) } &&   \label{null_vect}\\
- \frac{(x-1) x \partial_x}{(x-z) 
(z-1) z} + \frac{(1-2 z) \partial_z}{(z-1) z} + \frac{1}{b^2} \partial^2_z  &&\hspace{-0.7cm}\Big]\Psi_5(z,x) = 0 \,, \nn
\ea
where
\be
\Psi_5(z,x) = \langle  V_{\ha}(\infty) V_{(2,1)}(z) V_{\hb}(1) V_{\hc}(x) V_{\hd}(0) \rangle  \,,  \label{four_point}
\ee
from spherical variables $(x,z)$ to elliptic variables $(\tau, u)$, we will need the following derivatives:
\be  \label{dudz}
\frac{\partial u}{\partial z} = \frac{1}{\wp'} \frac{\partial \wp }{\partial z} = \frac{1}{2 \sqrt{e_1-e_3} \sqrt{ z (z-1) (z-x)}}  \,,
\ee
\be   \label{dtaudx}
\frac{\partial \tau}{\partial x} = \frac{ 4 \pi i}{(e_{1}-e_{3}) 4x (x-1)}  \,,
\ee
\be
\dudx = \frac{\partial_u \log \vartheta_0(u)}{2 (e_{1}-e_{3}) (1-x)x}  - \frac{1}{2(x-1)}\sqrt{\frac{z-1}{ (e_{1}-e_{3})  z(z-x)}}  \,.
\ee
The relations (\ref{dudz}) and (\ref{dtaudx}) are easily obtained from (\ref{diff_wp}) and (\ref{e_as_theta}). To compute  $\dudx$, we have followed references \cite{Manin,Takasaki:2000zd,Zabrodin:2011fk}.

The ansatz
\be
\Psi_5(x,z) = x^{-\frac{b^2}{4}} \left( (1-z) z  (z-x)\right)^{\frac{1}{4}+\frac{b^2}{2}} \vartheta_1(u)^{b^2} \psi(u,\tau)  \label{ansatz_rem_lin_div}
\ee 
yields a differential equation for $\psi(u,\tau)$ with vanishing linear derivative in $u$. We have here used the identities (\ref{wp_via_theta}), (\ref{one_prime_via_others}) and  (\ref{e_as_theta}) to express
\be
\vartheta_0(u) = \frac{\vartheta_1(u)}{x^{\frac{1}{4}} \,\sn(2 K u)} = \frac{\sqrt{z}}{x^{\frac{1}{4}} } \vartheta_1(u)\,,
\ee
which renders the prefactor in (\ref{ansatz_rem_lin_div}) more symmetric. 

To evaluate the $u$- and $\tau$-derivatives
 on the prefactor of $\psi(u,\tau)$ in the definition (\ref{ansatz_rem_lin_div}), we need the derivative
\ba
\dzdtau &=& - \frac{\partial x}{\partial \tau} \frac{\partial u}{\partial x} /\frac{\partial u}{\partial z} \\
&=& -\frac{1}{\pi} \left( i  (e_{1}-e_{3})  x (z-1) + \partial_u \log \vartheta_0 \sqrt{ (e_{1}-e_{3}) } \sqrt{z(z-1)(z-x)} \right) \,.  \nn
\ea
We can use the heat equation (\ref{heat}) and (\ref{wp_via_theta}) to express $\partial_u \log \vartheta_0$ in terms of $x$, $z$ and $\eta$.

Upon dividing the equation (\ref{null_vect}) by $\left( \frac{\partial u}{\partial z} \right)^2$, analyzing the pole and vanishing structure of the terms proportional 
to the weights $h_i$ fixes the form of these terms in elliptic variables to be proportional to
\be
\sum_{i=0}^{3} h_i ( \wp(u - \omega_i) - e_{i+2}) \,,
\ee
plus a $\tau$-dependent function. We
 have introduced $\omega_0=e_0=0$
 and indexed the half-periods modulo
$4$ for ease of notation.
The terms generated by our
ansatz (\ref{ansatz_rem_lin_div}) conspire to shift the
weights $h_i$ by a common term
\be
h_i \rightarrow h_i - \frac{b^4 + 2 b^2 +\frac{3}{4} }{4b^2} =: \hat{h}_i \,.
\ee
After a detailed calculation, we arrive at the differential equation
\be   \label{intermid_null}
\left(\partial_u^2 + 4 b^2 \sum_{i=0}^3 \hat{h}_i (\wp(u+ \omega_i) - e_{i+2}) \right) \psi =  -\left(f(\tau) + 4 \pi i b^2 \partial_\tau  \right)\psi  
\ee
with 
\be
f(\tau) = \frac{b^2}{3} \left( (e_3-e_1) (1 + 8 \hc) -(e_3-e_2) ( 2+ 3 b^2+16 \hc) - 6 \eta_1 (1+b^2)  \right) \,.
\ee
As the function $f(\tau)$ is $u$-independent, it can be absorbed in a
redefinition of the function $\psi$. Using the identities (\ref{e_as_theta}), we
determine the required rescaling to be
\be \psi(u,\tau) \rightarrow \vartheta_1'(0)^{-\frac{1}{3}(1+b^2)}
x^{-\frac{1}{12}(1+8 \hc)} (1-x)^{-\frac{1}{12}(1+3b^2+8 \hc)}
\psi(u,\tau) \,.  \ee 
Finally, we can also absorb the $u$-independent
term 
\be \sum_{i=1}^3 h_{i+2} \, e_{i}(\tau) \ee 
in
(\ref{intermid_null}) via a shift 
\be \psi(u,\tau) \rightarrow 
\prod_{i=1}^3 \left(\frac{ \eta
  }{\vartheta_{i+1}(0)} \right)^{4 h_{i+2}} \psi(u,\tau) \,.  
\ee 
We
have here used the identity (\ref{e_i_via_theta}). The differential equation in its
final form is thus given by 
\be \label{final_form} \left(\partial_u^2
  + 4 b^2 \sum_{i=0}^3 \hat{h}_i \wp(u+ \omega_i) \right) \psi(u,\tau)
= - 4 \pi i b^2 \partial_\tau \psi (u,\tau) \,, 
\ee 
where
$\psi(u,\tau)$ is related to the five-point function
(\ref{four_point}) via the rescaling
\be
\Psi_5(z,x) = \frac{\left[ (1-z) z  (z-x)\right]^{\frac{1}{4}+\frac{b^2}{2}} }{ \left[4 x(1-x) \right]^{\frac{1}{12}(1+3b^2+8 \hc)}}  \frac{\vartheta_1(u)^{b^2}}{\vartheta_1'(0)^{\frac{1}{3}(1+b^2)}} \prod_{i=1}^3 
\left(\frac{  \eta  }{\vartheta_{i+1}(0)} \right)^{4 h_{i+2}} \psi(u,\tau) \,.
\ee
We have chosen a overall normalization that is convenient in determining transformation
properties under channel duality.

\subsection{Properties of elliptic and modular functions}
\label{identities}
The Weierstrass $\wp$-function is an even function with a double pole at the origin. It satisfies the differential equation
\be   \label{diff_wp}
\wp'(v)^2 = 4 \prod_{i=1}^3 (\wp(v) - e_i) \,.
\ee
The $e_i$ are the half-periods of $\wp$. 
Since the function $\wp$ is even, we conclude that they are the images of the 2-torsion points of the torus under $\wp$,
\be  \label{half_periods}
e_i = \wp(\omega_i) \,.
\ee
Denoting the periods of $\wp$ as $\omega$ and $\omega'$, we enumerate the 2-torsion points as $\omega_1=\frac{\omega}{2}$, $\omega_2=\frac{\omega+\omega'}{2}$, $\omega_3=\frac{\omega'}{2}$. With regard to a rescaling of its periods, the Weierstrass $\wp$-function behaves as
\be
\wp(v \,| \,\omega, \omega') = \frac{1}{\omega^2} \wp \left( \frac{v}{\omega} \, \Big| \, 1, \frac{\omega'}{\omega} \right) \,.
\ee
When relating $\wp$ to theta functions below and in our computations in the body of this paper, we will normalize its periods to $1$ and $\tau$ to simplify the formulae.

The Jacobi sine function provides a two-to-one odd map from the torus to the sphere. It relates to the Weierstrass function $\wp$ as follows,
\be \label{sn_from_wp}
\sn^2(\sqrt{e_1-e_3}\, v) =  \frac{e_1-e_3}{\wp(v)-e_3}  \,.
\ee
Defining $w =\sqrt{e_1-e_3}\, v$ and introducing the standard notation
$K = \sqrt{e_1-e_3}\, \omega_1$, $iK'=\sqrt{e_1-e_3}\, \omega_3$,
the sine function $\sn(w)$ has periodicities $4 K$ and $2 i K'$,
and moreover satisfies $\sn(w+2K) =
-\sn(w) $. The scaling of the argument of $\sn(w)$ in (\ref{sn_from_wp}) is to render 
its periods independent of the choice of periods of $\wp$. The Jacobi sine function obeys the differential equation
\be
\sn'(w)^2 = (1- \sn(w)^2)(1-k^2 \sn(w)^2)
\ee
with $k^2 =x$, and can hence be locally inverted via
\be  \label{inv_sn}
\sn^{-1}(z) = \int_0^z \frac{dy}{\sqrt{(1-y^2)(1-k^2 y^2)}}  \,.
\ee
The sine function with shifted argument satisfies the relations
\ba
\sn(w+K) &=& \sqrt{\frac{1-\sn^2(w)}{1- k^2 \,\sn^2(w)}} \,, \\
\sn(w+iK') &=& \frac{1}{k \, \sn(w)} \,, \\
\sn(w+K+iK') &=& \frac{1}{k \, \sn(w+K)} \,. 
\ea
These relations can be invoked to derive expressions for 
the shifted Weierstrass function $\wp(v+\omega_i)$ in terms of $\wp(v)$ and the half-periods.

Theta functions $\vartheta_i( u | \tau) = \vartheta_i(u)$, $i=0,
\ldots, 3$,
are quasi-periodic functions associated to the lattice
spanned by the lattice vectors $1$ and $\tau$. Both the Weierstrass
function $\wp$ and its half-periods can be expressed in terms of the
theta functions $\vartheta_i$ and their $\tau$-derivatives, related to the $u$-derivative via the heat equation 
\be \label{heat} 4 \pi i
\,\partial_\tau \vartheta_i - \partial_u^2 \vartheta_i =0 \,.  
\ee 

Thus, the function $\wp$ can be expressed via the theta functions
as 
\be
\wp (u) = - \partial_u\log \vartheta_1 - 2 \eta_1
\ee
and
\be \label{wp_via_theta} \sqrt{\wp(u)-e_i} = \frac{\vartheta_1'(0)}{\vartheta_{i+1}(0)}
\frac{\vartheta_{i+1}(u)}{\vartheta_1(u)} \, ,  
\ee
where indices are to be understood modulo 4.
We have here introduced the notation
\be \eta_1 = - \frac{2 \pi i }{3} \partial_\tau \! \log\vartheta_1'(0) = -2\pi i \partial_\tau \log \eta\,, \nonumber 
\ee 
with $\eta$ the Dedekind eta
function that satisfies 
\be 
\eta = \left(\frac{\vartheta_1'(0)}{2 \pi    } \right)^{\frac{1}{3}} \,. 
\ee
Expanding the right-hand side of equation (\ref{wp_via_theta}) in $u$ and recalling that the Weierstrass function expanded around the origin has no constant term yields the expressions
\be \label{e_i_via_theta} 
e_i = -  4 \pi i \, \partial_\tau \! \log \vartheta_{i+1}(0) - 2 \eta_1 
\ee
for the half-periods in terms of theta functions. 
Evaluating equation (\ref{wp_via_theta}) at the 2-torsion points and invoking
\be \label{one_prime_via_others} 
\vartheta_1'(u) = \pi \,\vartheta_0(u) \vartheta_2(u) \vartheta_3(u) 
\ee 
gives rise respectively to the first equality in the following three identities:
\ba
e_1 - e_2 &=& \pi^2 \vartheta_0^4(0) = 4 \pi i \partial_\tau \log \frac{\vartheta_3(0)}{\vartheta_2(0)} \,, \label{e_as_theta} \\
e_1 - e_3 &=& \pi^2 \vartheta_3^4(0) = 4 \pi i \partial_\tau \log \frac{\vartheta_0(0)}{\vartheta_2(0)} \,,  \nn \\
e_2 - e_3 &=& \pi^2 \vartheta_2^4(0) = 4 \pi i \partial_\tau \log
\frac{\vartheta_0(0)}{\vartheta_3(0)} \,.  \nn 
\ea 
The second then follow after recourse to equation (\ref{e_i_via_theta}).

\section{The monodromies of shifted Weierstrass integrands}
\label{Weierstrassmonodromies}
In imposing the monodromy condition on our semi-classical conformal block,
we need to compute the following types of integrals:
\begin{eqnarray}
K(n_0,n_1,n_2,n_3) &=& \oint_\alpha \wp (z)^{n_0} \wp (z+\omega_1)^{n_1} \wp(z+\omega_2)^{n_2} \wp(z+\omega_3)^{n_3}.
\end{eqnarray}
This is a class of integrals generalizing those reviewed in
\cite{KashaniPoor:2012wb} -- we will freely use the results of that
reference.  In fact, we also encounter
 integrands containing
an even number of derivatives of the Weierstrass function evaluated at half-periods, but
these can be eliminated using the formulae
\begin{eqnarray}
 (\wp'(z+\omega_i))^2 &=& 4 \wp(z+\omega_i)^2 - g_2 \wp(z+\omega_i) -g_3 
\nonumber 
\end{eqnarray}
and
\begin{eqnarray}
\wp'(z+\omega_i) \wp'(z+\omega_j) &=& 3 ( - \wp^2(z+\omega_i) \wp(z+ \omega_j) -\wp(z+\omega_i) \wp^2(z+ \omega_j)
\nonumber \\
& & 
+ \frac{1}{12} g_2 (\wp(z+\omega_i) +\wp(z+ \omega_j))\nonumber \\
& &  + e_{ij} (\wp(z+\omega_i)^2+ \wp^2(z+ \omega_j)
- \frac{1}{6} g_2)) \, ,
\end{eqnarray}
where $e_{ij} = \wp ( \omega_i + \omega_j)$. Integrands with an odd number of derivatives vanish,
 as can be seen
from the parity of the integrand  and the symmetry of the
 integration domain.
Finally, the integrals $K(n_0,n_1,n_2,n_3)$ can be computed using the identity
\begin{eqnarray}
\wp(z+\omega_i) \wp(z) &=& e_{i} (\wp(z+\omega_i) + \wp(z)) + e_i^2+\prod_{k \neq i} e_k  \,.
\end{eqnarray}
 Indeed, by invoking this identity,
 the total order $n_0+n_1+n_2+n_3$ of the integrand
of $K(n_0,n_1,n_2,n_3)$ can be lowered until
 the integrand is a power of a single shifted
Weierstrass function, which can then be computed using the results of \cite{KashaniPoor:2012wb}.
In other words, the recursion relation
\begin{eqnarray}
K(n_0,n_1,n_2,n_3) &=&  e_1 (K(n_0,n_1-1,n_2,n_2) + K(n_0-1,n_1,n_2,n_3) 
\nonumber \\
& & 
+ (e_1^2 + e_2 e_3) 
K[n_0-1,n_1-1,n_2,n_3))
\end{eqnarray}
and its close cousins are sufficient to reduce the evaluation of monodromies of products of shifted Weierstrass functions to known integrals.

\section{The group action} \label{group}
\begin{figure}[htb!]
\centering%
\includegraphics{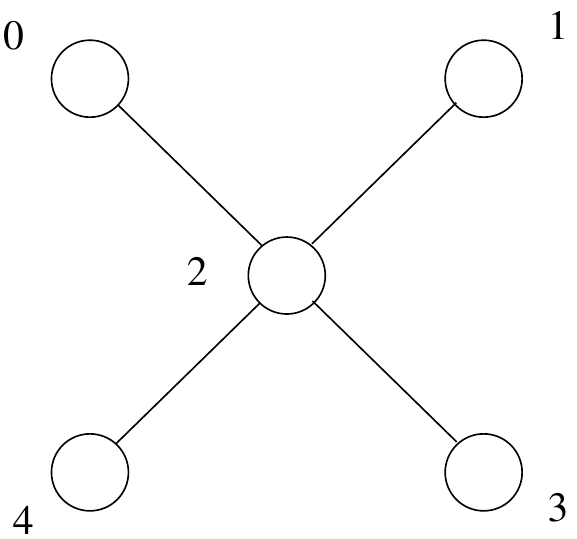}
\caption{The Dynkin diagram of affine $D_4^{(1)}$. }
\label{affineso8}
\end{figure}

The Dynkin diagram of affine $D_4^{(1)}$
 is depicted in figure \ref{affineso8}.
If we delete the affine node labeled by zero, we obtain the Dynkin
diagram of $D_4$. The outer automorphisms of $D_4$, i.e. triality,
act by exchanging the external nodes of the $D_4$ Dynkin diagram.
With the basis $\{\epsilon_i \}$ of root space introduced in section
\ref{rev_nf4}, we have
\begin{eqnarray}
\alpha_0^\lor &=& \delta - (\epsilon_1 + \epsilon_2)  \,,
\nonumber \\
\alpha_1 &=& \epsilon_1 - \epsilon_2 \,,
\nonumber \\
\alpha_2 &=& \epsilon_2 - \epsilon_3 \,,
\nonumber \\
\alpha_3 &=& \epsilon_3 - \epsilon_4 \,,
\nonumber \\
\alpha_4 &=& \epsilon_3 + \epsilon_4 \,.
\end{eqnarray}
The group $SL(2,\mathbb{Z})$ is the semi-direct product $SL(2,\mathbb{Z}) = \Gamma(2) \rtimes S_3$.
If we write elements of the set $\Gamma(2) \times S_3$ as pairs $(\gamma,p)$, then the group structure
of $SL(2,\mathbb{Z})$ 
is given by $(\gamma_1,p_1)\cdot (\gamma_2,p_2) = (\gamma_1 a_{p_1} (\gamma_2), p_1 p_2)$
where $a_{p_1}$ are the automorphisms of $\Gamma(2)$ induced by conjugation of $\Gamma(2)$
by the elements $S,ST,STS,STST,STSTS,T$ of $SL(2,\mathbb{Z})$.

The full symmetry group of the theory is $( W(D_4) \times
 \Gamma(2)) \rtimes S_3$, 
where $S_3$ acts as outer automorphisms on $SO(8)$ and by automorphisms
inner to $SL(2,\mathbb{Z})$ on $\Gamma(2)$. The composition in the symmetry group is
 $(g_1,\gamma_1,p_1)\cdot (g_2,\gamma_2,p_2) = (g_1 \tau_{p_1} (g_2) ,
\gamma_1 a_{p_1} (\gamma_2), p_1 p_2)$ where $\tau_{p_1}$ is the outer automorphism of $SO(8)$ associated
to the permutation element $p_1$ of $S_3$. 

Specifically, the $S$ transformation permutes the vector and the
spinor representation in the spectrum, while the $T$ transformation
permutes the spinor and the conjugate spinor representation
\cite{Seiberg:1994aj}. With the fundamental weights 
$\mu_1,
\mu_3,\mu_4$, dual to the respective simple roots, corresponding to
the vector, spinor, and conjugate spinor respectively, we conclude
that $p_S$ permutes $\alpha_1$ and $\alpha_3$, while $p_T$ permutes
$\alpha_3$ and $\alpha_4$. It can be checked that this convention is consistent
with our convention for the cross ratio $x$ as a function of the modular parameter
$\tau$, and the action of the modular group on the masses $M_i$.

\bibliography{cft}
\bibliographystyle{utcaps}

\end{document}